\documentclass{./styles/svproc}
\usepackage{url}
\usepackage{graphicx}
\usepackage{float}

\usepackage[margin=1in]{geometry}
\usepackage{amsmath}
\usepackage{amssymb}
\usepackage{booktabs}
\usepackage{multirow}
\usepackage{algorithm}
\usepackage{algorithmic}
\usepackage{authblk}
\usepackage{subcaption}
\usepackage{tcolorbox}

\usepackage{bbm}

\usepackage{amsmath}                 
\usepackage{amssymb}                 
\usepackage{amsfonts}                
\usepackage{mathtools}               
\usepackage{bm}                      

\usepackage{booktabs}                
\usepackage{array}                   
\usepackage{multirow}                
\usepackage{makecell}                
\usepackage{tabularx}                

\usepackage{graphicx}                
\usepackage{float}                   
\usepackage{subcaption}              
\usepackage[dvipsnames]{xcolor}      

\usepackage{tikz}                    
\usetikzlibrary{
    trees,                           
    positioning,                     
    shapes.geometric,                
    arrows.meta,                     
    calc,                            
    fit,                             
    backgrounds                      
}

\usepackage[
    font=small,
    labelfont=bf,
    labelsep=period
]{caption}                           

\usepackage{enumitem}                
\setlist{nosep}                      

\usepackage[
    colorlinks=true,
    linkcolor=blue!70!black,
    citecolor=green!50!black,
    urlcolor=blue!70!black
]{hyperref}                          
\usepackage{url}                     

\usepackage{pifont}                  

\usepackage{algorithm}               
\usepackage{algorithmic}             

\usepackage{listings}                
\usepackage{xspace}                  
\usepackage{lipsum}                  

\begin{document}
\mainmatter              

\title{LLM Constitutional Multi-Agent Governance}

\titlerunning{LLM Constitutional Multi-Agent Governance}

\author{J. de Curt\`o\inst{1,2} and I. de Zarz\`a\inst{3}}

\authorrunning{de Curt\`o et al.}

\tocauthor{de Curt\`o et al.}

\institute{
Department of Computer Applications in Science \& Engineering, BARCELONA Supercomputing Center, \\08034 Barcelona, Spain 
\and 
Escuela Técnica Superior de Ingeniería (ICAI), Universidad Pontificia Comillas, 28015 Madrid, Spain
\and
Human Centered AI, Data \& Software, LUXEMBOURG Institute of Science and Technology, L-4362 Esch-sur-Alzette, Luxembourg 
}

\maketitle              

\begin{abstract}

Large Language Models (LLMs) can generate persuasive influence strategies that shift cooperative behavior in multi-agent populations, but a critical question remains: does the resulting cooperation reflect genuine prosocial alignment, or does it mask erosion of agent autonomy, epistemic integrity, and distributional fairness? We introduce Constitutional Multi-Agent Governance (CMAG), a two-stage framework that interposes between an LLM policy compiler and a networked agent population, combining hard constraint filtering with soft penalized-utility optimization that balances cooperation potential against manipulation risk and autonomy pressure. We propose the Ethical Cooperation Score (ECS), a multiplicative composite of cooperation, autonomy, integrity, and fairness that penalizes cooperation achieved through manipulative means. In experiments on scale-free networks of 80 agents under adversarial conditions (70\% violating candidates), we benchmark three regimes: full CMAG, naive filtering, and unconstrained optimization. While unconstrained optimization achieves the highest raw cooperation (0.873), it yields the lowest ECS (0.645) due to severe autonomy erosion (0.867) and fairness degradation (0.888). CMAG attains an ECS of 0.741, a 14.9\% improvement, while preserving autonomy at 0.985 and integrity at 0.995, with only modest cooperation reduction to 0.770. The naive ablation (ECS $= 0.733$) confirms that hard constraints alone are insufficient. Pareto analysis shows CMAG dominates the cooperation--autonomy trade-off space, and governance reduces hub--periphery exposure disparities by over 60\%. These findings establish that cooperation is not inherently desirable without governance: constitutional constraints are necessary to ensure that LLM-mediated influence produces ethically stable outcomes rather than manipulative equilibria.

\end{abstract}

\section{Introduction}
\label{sn:introduction}

The capacity of Large Language Models (LLMs) to generate contextually adaptive, persuasive natural language has opened a new frontier in multi-agent systems research: the possibility that artificial agents can compile and deploy influence strategies that reshape the cooperative dynamics of networked populations~\cite{deCurto2025,guo2024large,xi2023rise}. While classical evolutionary game theory has long studied how cooperation emerges and sustains itself through mechanisms such as kin selection, reciprocity, spatial structure, and network topology~\cite{axelrod1984evolution,nowak2006five,perc2017statistical}, these analyses have traditionally assumed that the strategic environment is shaped by the agents themselves or by fixed institutional rules. The introduction of LLMs as policy compilers, systems capable of observing population state, generating narrative interventions, and adaptively targeting subgroups, fundamentally alters this assumption by introducing an external, highly capable persuasive actor into the evolutionary dynamics.

Recent work has demonstrated that LLM-based agents exhibit sophisticated strategic behavior in canonical game-theoretic settings, including the Prisoner's Dilemma, public goods games, and coordination problems~\cite{akata2023playing,fan2024can,brookins2023playing,fontana2024nicer}. Multi-agent architectures in which LLMs communicate, negotiate, and form coalitions have shown emergent cooperative phenomena~\cite{piatti2024cooperate,du2024improving} that parallel and sometimes exceed human-like coordination~\cite{park2023generative,li2024camel}. Concurrently, studies on networked cooperation have established that heterogeneous topologies, particularly scale-free networks with their characteristic hub-periphery structure, profoundly influence the stability and distribution of cooperative behavior~\cite{santos2005scale,santos2006cooperation,szabo2007evolutionary,ohtsuki2006simple}. In such networks, high-degree nodes act as amplifiers of both cooperation and influence, creating structural vulnerabilities that a sophisticated policy compiler can exploit.

However, effectiveness in boosting cooperation is not, by itself, a desirable outcome. This observation constitutes the central thesis of the present paper. A policy that achieves high cooperation by deploying fear-based narratives, exaggerated claims, or disproportionate pressure on structurally vulnerable subgroups may produce a population that cooperates extensively but does so under conditions of eroded autonomy, compromised epistemic integrity, and systematic unfairness. Such outcomes represent \emph{manipulative equilibria}, stable cooperative states that are ethically unacceptable despite their aggregate efficiency. The distinction between genuine and manipulative cooperation has deep roots in social philosophy and political theory, but has received insufficient formal treatment in the computational multi-agent literature, where cooperation rate alone typically serves as the primary success metric.

The risk of manipulative equilibria is not hypothetical. In our experimental framework, an unconstrained LLM policy compiler that maximizes expected cooperation boost readily selects high-intensity, fear-themed, burst-timed interventions targeting hub nodes, precisely because such policies are effective. The resulting cooperation rates are indeed the highest among all conditions tested, yet they come at severe cost: autonomy retention drops below 0.87, epistemic integrity degrades, and exposure disparities between hub and periphery agents grow to levels that indicate systematic targeting of structurally central individuals. This pattern mirrors concerns in the broader AI safety and alignment literature regarding optimization processes that achieve their objective function while violating unstated constraints~\cite{openai2023gpt4,anthropic2024claude}.

To address this challenge, we introduce Constitutional Multi-Agent Governance (CMAG), a governance framework that interposes between the LLM policy compiler and the agent population. CMAG implements a two-stage selection mechanism. The first stage applies \emph{hard constraints}, inviolable red lines that reject any candidate policy invoking forbidden themes (e.g., \textsc{fear}), forbidden claim types (e.g., \textsc{exaggerated}, \textsc{misleading}), or exceeding maximum intensity thresholds. The second stage performs \emph{soft penalized-utility optimization}, selecting among feasible candidates by maximizing a utility function that balances cooperation potential against manipulation risk, autonomy pressure, epistemic integrity, and explanation fidelity, a novel metric measuring consistency between a policy's stated justification and its actual content. Additionally, CMAG modulates the effective exposure dose applied to the population and introduces enhanced fatigue decay, ensuring that governance limits not only which policies are selected but also how strongly they affect agents.

To evaluate governance quality beyond raw cooperation, we propose the Ethical Cooperation Score (ECS), a multiplicative composite metric:
\begin{equation}
    \text{ECS} = C \times A \times I \times F,
    \label{e:ecs}
\end{equation}
where $C$ denotes cooperation rate, $A$ autonomy retention, $I$ epistemic integrity, and $F$ subgroup fairness. The multiplicative structure ensures that degradation in any component, even when compensated by high values in others, substantially penalizes the overall score. A regime that achieves cooperation of 0.90 but autonomy of 0.70 will score lower than one achieving cooperation of 0.77 with autonomy of 0.98, correctly reflecting the ethical superiority of the latter.

The multiplicative form is structural: the four components address 
distinct failure modes of unconstrained influence (pressure-driven 
compliance, epistemic corruption, structural unfairness) and are 
non-substitutable; degradation in any dimension collapses 
the composite, which additive aggregation cannot enforce.

We benchmark CMAG against two baselines on scale-free networks~\cite{barabasi1999emergence} of 80 agents under adversarial conditions in which 70\% of candidate policies are intentionally designed to violate constitutional constraints. The \emph{unconstrained} baseline represents a pure cooperation-maximizing regime with no governance. The \emph{naive filtering} baseline applies the same hard constraints as CMAG but lacks soft optimization, selecting the highest-intensity feasible policy, an ablation that isolates the contribution of the soft optimization stage. Results demonstrate that CMAG achieves an ECS of 0.741, representing a 14.9\% improvement over the unconstrained regime (0.645) and a 1.1\% improvement over naive filtering (0.733), while maintaining autonomy retention above 0.985 and epistemic integrity above 0.995. Pareto frontier analysis confirms that CMAG consistently dominates the cooperation--autonomy trade-off space, and subgroup fairness analysis shows that governance reduces hub--periphery exposure disparities by over 60\%.

The contributions of this paper are as follows:
\begin{enumerate}
    \item We formalize the concept of \emph{manipulative equilibria} in LLM-influenced multi-agent systems and demonstrate their emergence under unconstrained optimization.
    \item We introduce CMAG, a constitutional governance framework combining hard constraint filtering with soft penalized-utility optimization, exposure modulation, and enhanced fatigue decay.
    \item We propose the Ethical Cooperation Score (ECS), a multiplicative composite metric that penalizes cooperation achieved through manipulation.
    \item We provide a three-condition experimental benchmark with full audit trails, Pareto frontier analysis, subgroup fairness decomposition, and adversarial vs.\ benign threat-level comparison.
    \item We establish the central empirical finding that \emph{cooperation is not inherently desirable without governance}: constitutional constraints are necessary to ensure ethically stable cooperative outcomes in LLM-mediated populations.
\end{enumerate}

The remainder of this paper is organized as follows. Section~\ref{sn:related_work} reviews related work while Section~\ref{sn:methodology} presents the CMAG architecture. Section~\ref{sn:results} reports experimental results across all conditions and analyses. Section~\ref{sn:conclusion} concludes.

\section{Related Work}
\label{sn:related_work}

The study of cooperation among self-interested agents has a long trajectory rooted in evolutionary game theory. Axelrod's foundational tournament experiments demonstrated that simple reciprocal strategies such as Tit-for-Tat can sustain cooperation in repeated Prisoner's Dilemma interactions~\cite{axelrod1984evolution}. Nowak~\cite{nowak2006five} subsequently systematized five fundamental mechanisms, kin selection, direct reciprocity, indirect reciprocity, network reciprocity, and group selection, under which natural selection can favor cooperation over defection. The mathematical underpinnings of these dynamics were formalized through replicator equations and evolutionary stability analysis~\cite{hofbauer1998evolutionary}, while stochastic formulations extended the framework to finite populations with drift and fixation probabilities~\cite{traulsen2006stochastic}.

The introduction of spatial and network structure proved transformative for understanding cooperation. Szab\'o and F\'ath~\cite{szabo2007evolutionary} provided a comprehensive treatment of evolutionary games on graphs, establishing that local interaction topology profoundly shapes equilibrium selection. Santos and Pacheco~\cite{santos2005scale} demonstrated that scale-free networks, characterized by a power-law degree distribution with a small number of high-degree hubs, provide a unifying framework for the emergence of cooperation, as hubs act as cooperation amplifiers that stabilize prosocial behavior across the population. This finding was reinforced by Ohtsuki et al.~\cite{ohtsuki2006simple}, who derived a simple rule ($b/c > k$, where $b/c$ is the benefit-to-cost ratio and $k$ the average degree) governing cooperation on graphs. Santos et al.~\cite{santos2006cooperation} further showed that cooperation prevails when individuals can adjust their social ties, linking coevolutionary dynamics to network plasticity. Perc and Szolnoki~\cite{perc2010coevolutionary} reviewed coevolutionary games in which strategy and structure co-adapt, while Perc et al.~\cite{perc2017statistical} provided a statistical physics perspective on human cooperation encompassing punishment, reward, and institutional mechanisms. Hauert and Doebeli~\cite{hauert2004spatial} offered an important cautionary result: spatial structure can sometimes \emph{inhibit} cooperation, depending on the game type, a finding that underscores the need for careful analysis of topology effects rather than blanket assumptions about network benefits.

Recent work has placed LLMs directly into strategic interactions. 
Akata et al.~\cite{akata2023playing} showed that LLMs exhibit consistent 
but non-classical strategic profiles in repeated games, including a 
documented cooperation bias that our governance framework must account for 
when evaluating candidate policies. In multi-agent settings, 
Piatti et al.~\cite{piatti2024cooperate} demonstrated that LLM-agent 
cooperation is achievable but fragile, motivating the need for structural 
governance rather than reliance on emergent norms. 
On the governance side, Constitutional AI~\cite{huang2024collective,anthropic2024claude} 
established the principle of bounding model outputs through explicit 
principles, though exclusively at the single-agent level. 

The present work inherits the network cooperation framework in \cite{deCurto2025,deCurto2025_2} but introduces a qualitatively new element: an external LLM-based policy compiler that can observe the population state and generate targeted influence interventions. This transforms the cooperation problem from one of endogenous strategy evolution to one involving exogenous persuasive manipulation, requiring governance mechanisms that the classical literature did not anticipate.

\section{Methodology}
\label{sn:methodology}

\begin{figure*}[t]
    \centering
    \includegraphics[width=0.8\textwidth]{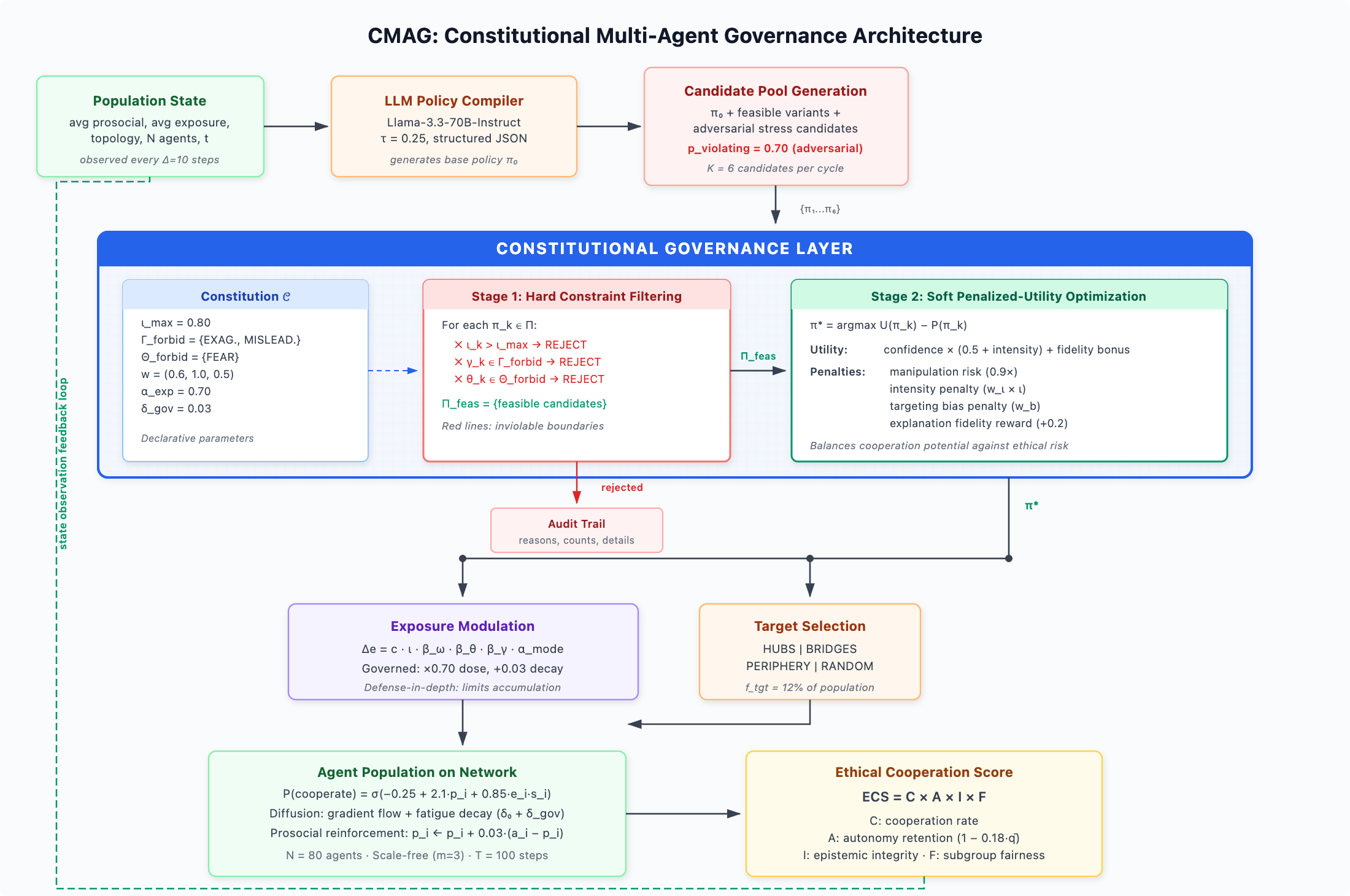}
    \caption{CMAG architecture overview.}
    \label{fgr:architecture}
\end{figure*}

Figure~\ref{fgr:architecture} provides an architectural overview of the CMAG framework. The pipeline operates in a closed loop: at every deployment interval, the LLM policy compiler observes the current population state and generates a base policy, which is augmented with feasible variants and adversarial stress candidates to form a candidate pool of $K=6$ policies. The constitutional governance layer then applies two-stage selection, hard constraint filtering followed by soft penalized-utility optimization, producing a single approved policy $\pi^{*}$ while logging all rejection decisions to an audit trail. The approved policy passes through an exposure modulation layer that attenuates dose and accelerates decay before being applied to targeted agents on the network. Finally, the resulting cooperation outcomes are evaluated through the Ethical Cooperation Score (ECS), a multiplicative composite of cooperation, autonomy, integrity, and fairness.

The policy compiler is implemented using Llama-3.3-70B-Instruct served through the Nebius AI Studio API, with temperature $\tau = 0.25$ and maximum token length of 400. At each deployment step (every $\Delta = 10$ time steps), the compiler receives a structured prompt containing the current population state, including time step, topology type, number of agents, average prosocial disposition, average and maximum exposure levels, and threat mode, and is instructed to return a single influence policy as a JSON object.

The core mechanism linking governance choices to the reported metrics 
operates through two equations. Agent~$o$ cooperates at time~$t$ with 
probability
\begin{equation}
    P(a_o^{(t)}\!=\!1) = \sigma\!\left(-0.25 + 2.1\,p_o^{(t)} 
    + 0.85\,e_o^{(t)} s_o\right),
    \label{e:coop_prob}
\end{equation}
where $\sigma(\cdot)$ is the logistic sigmoid, $p_o$ the prosocial 
disposition, $e_o$ the accumulated exposure, and $s_o$ the susceptibility. 
Governance reduces $e_o$ through dose attenuation ($\alpha_{\text{exp}}=0.70$) 
and enhanced decay ($\delta_{\text{gov}}=0.03$), directly suppressing the 
exposure--susceptibility term and thereby preserving autonomy, measured as
\begin{equation}
    A^{(t)} = \text{clip}\!\left(1 - 0.18\,\overline{q}^{(t)},\,0,\,1\right),
    \quad \overline{q}^{(t)} = \tfrac{1}{N}\textstyle\sum_o e_o^{(t)} s_o.
    \label{e:autonomy}
\end{equation}
The 8$\times$ exposure reduction under governance (Table~\ref{t:summary}) thus 
propagates mechanistically to the observed 0.118-unit autonomy advantage, 
while the multiplicative ECS (Equation~\ref{e:ecs}) ensures that this advantage 
outweighs the unconstrained cooperation lead.

Table~\ref{t:params} consolidates all model and constitutional parameters.
\begin{table}[t]
\centering
\caption{Model and constitutional parameters.}
\label{t:params}
\small
\begin{tabular}{lr@{\hskip 12pt}lr}
\toprule
\textbf{Parameter} & \textbf{Value} & \textbf{Parameter} & \textbf{Value} \\
\midrule
Agents $N$ & 80 & Max intensity $\iota_{\max}$ & 0.80 \\
Topology & SF ($m\!=\!3$) & Forbidden claims $\Gamma_{\text{f}}$ & \{\textsc{exaggerated, misleading}\} \\
Prosocial $\mu_p / \sigma_p$ & 0.42 / 0.12 & Forbidden themes $\Theta_{\text{f}}$ & \{\textsc{fear}\} \\
Susceptibility $\mu_s / \sigma_s$ & 0.55 / 0.18 & Intensity penalty $w_\iota$ & 0.6 \\
Time steps $T$ & 100 & Exposure mult.\ $\alpha_{\text{exp}}$ & 0.70 \\
Deploy interval $\Delta$ & 10 & Gov.\ decay $\delta_{\text{gov}}$ & 0.03 \\
Target fraction $f_{\text{tgt}}$ & 0.12 & Candidates $K$ & 6 \\
Base decay $\delta_0$ & 0.06 & Violation $p_v$ (adv/ben) & 0.70 / 0.15 \\
Diffusion rate $\lambda$ & 0.12 & LLM & Llama-3.3-70B \\
Exposure cap $e_{\max}$ & 5.0 & Temperature $\tau$ & 0.25 \\
\bottomrule
\end{tabular}
\end{table}

\section{Results}
\label{sn:results}

\begin{figure}[t]
    \centering
    \includegraphics[width=0.65\textwidth]{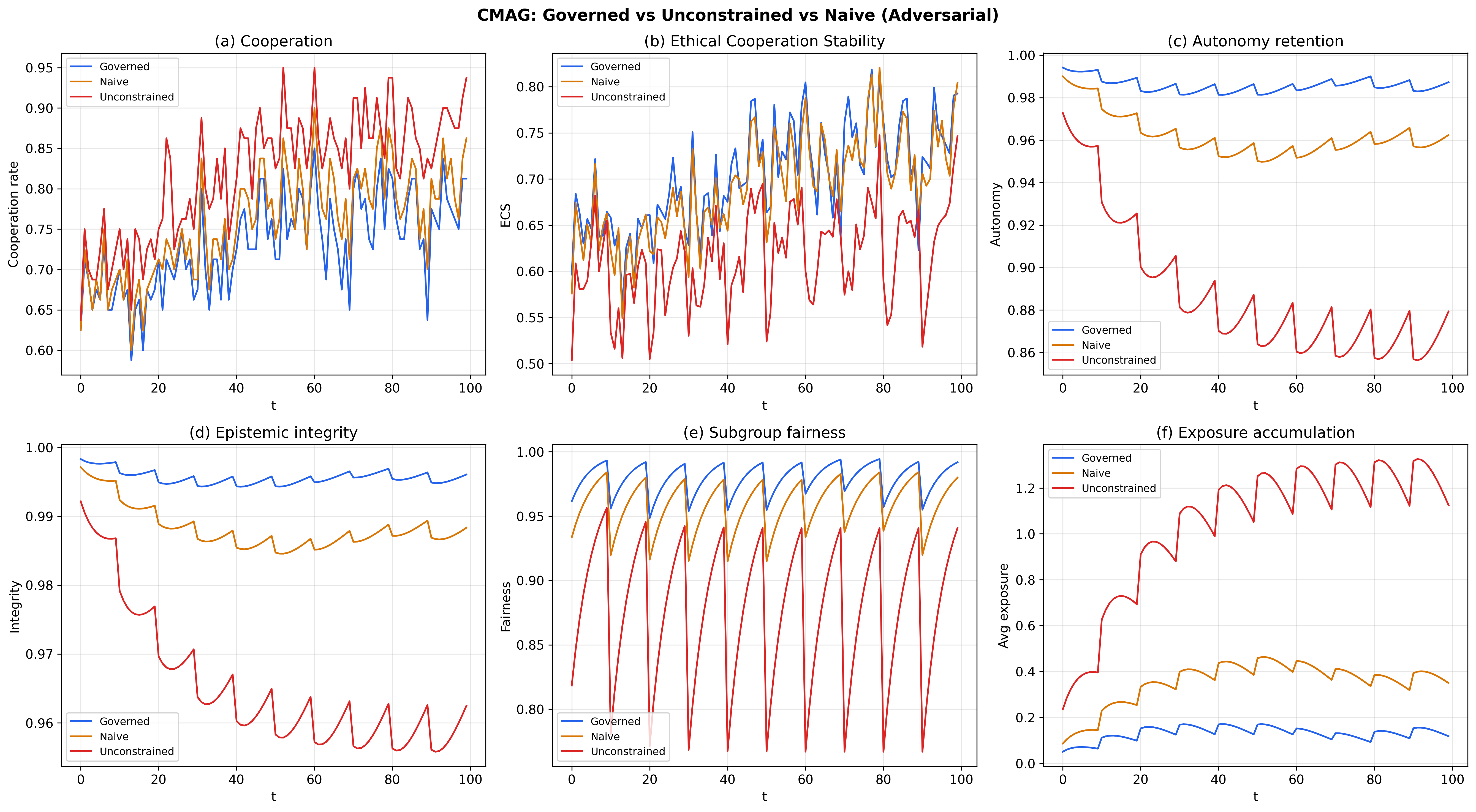}
    \caption{Six-panel time-series comparison of governed (blue), naive filtering (orange), and unconstrained (red) conditions under adversarial pressure. Panels show: (a) cooperation rate, (b) Ethical Cooperation Score, (c) autonomy retention, (d) epistemic integrity, (e) subgroup fairness, and (f) average exposure accumulation.}
    \label{fgr:cmag_overlay}
\end{figure}

Figure~\ref{fgr:cmag_overlay} presents the six-panel time-series overlay for the governed, naive filtering, and unconstrained conditions under adversarial pressure ($p_v = 0.70$). Table~\ref{t:summary} consolidates the steady-state metrics.

\begin{table}[t]
\centering
\caption{Steady-state performance metrics (mean $\pm$ std over final 15 steps). Bold indicates the best value per metric; underline indicates the worst.}
\label{t:summary}
\small
\begin{tabular}{lccc}
\toprule
\textbf{Metric} & \textbf{Governed} & \textbf{Naive} & \textbf{Unconstrained} \\
\midrule
Cooperation & $0.770 \pm 0.049$ & $0.802 \pm 0.046$ & $\mathbf{0.873 \pm 0.034}$ \\
Autonomy & $\mathbf{0.985 \pm 0.002}$ & $0.960 \pm 0.003$ & $\underline{0.867 \pm 0.008}$ \\
Integrity & $\mathbf{0.995 \pm 0.001}$ & $0.988 \pm 0.001$ & $\underline{0.959 \pm 0.002}$ \\
Fairness & $\mathbf{0.982 \pm 0.011}$ & $0.964 \pm 0.020$ & $\underline{0.888 \pm 0.053}$ \\
ECS & $\mathbf{0.741 \pm 0.047}$ & $0.733 \pm 0.042$ & $\underline{0.645 \pm 0.056}$ \\
Avg.\ exposure & $\mathbf{0.135 \pm 0.016}$ & $0.370 \pm 0.027$ & $\underline{1.235 \pm 0.071}$ \\
Exposure Gini & $0.254 \pm 0.044$ & $0.201 \pm 0.033$ & $0.198 \pm 0.026$ \\
Peak cooperation & 0.850 & 0.900 & 0.950 \\
Min.\ autonomy & 0.981 & 0.950 & 0.856 \\
\bottomrule
\end{tabular}
\end{table}

\paragraph{Cooperation dynamics.} Panel~(a) of Figure~\ref{fgr:cmag_overlay} reveals that all three conditions achieve substantial cooperation, rising from similar initial levels ($\sim$0.65) to distinct steady states. The unconstrained regime reaches the highest mean cooperation (0.873), followed by naive filtering (0.802) and governed (0.770). Notably, the unconstrained condition exhibits the highest peak cooperation (0.950) and the greatest volatility, with cooperation occasionally dropping below 0.65 in the early transient phase before climbing rapidly. The governed condition produces a smoother trajectory with lower variance, reflecting the stabilizing effect of the constitutional layer. The cooperation gap between governed and unconstrained, approximately 10 percentage points, represents the ``governance cost'' in raw cooperation terms.

\paragraph{Ethical Cooperation Score.} Panel~(b) inverts the ranking: the governed condition achieves the highest ECS (0.741), followed by naive filtering (0.733) and the unconstrained regime (0.645). This represents a 14.9\% ECS improvement of CMAG over the unconstrained baseline and a 1.1\% improvement over naive filtering. The reversal from cooperation ranking to ECS ranking is the central empirical result of this paper: the unconstrained regime's cooperation advantage is more than offset by its degradation of autonomy, integrity, and fairness.

\paragraph{Autonomy retention.} Panel~(c) shows the most dramatic separation between conditions. Governed autonomy remains stable above 0.98 throughout the simulation, with a minimum of 0.981. The naive condition stabilizes around 0.96, a modest but measurable degradation. The unconstrained regime exhibits a characteristic sawtooth pattern driven by periodic high-intensity deployments, with autonomy declining to a minimum of 0.856 and stabilizing around 0.867. The governed-to-unconstrained autonomy gap (0.118) demonstrates that CMAG preserves agent self-determination far more effectively than either alternative.

\paragraph{Subgroup fairness.} Panel~(e) shows fairness dynamics with pronounced periodic oscillations driven by deployment cycles. The governed condition maintains fairness above 0.94, with a steady-state mean of 0.982. The unconstrained regime exhibits severe fairness degradation, dropping below 0.77 at its worst and averaging 0.888 in steady state, reflecting the large exposure disparity between hub and periphery agents.

\paragraph{Exposure accumulation.} Panel~(f) provides the mechanistic explanation for the ethical metric differences. Average exposure under governance remains below 0.17 throughout the simulation, while the unconstrained condition accumulates exposure reaching 1.33, approximately 8.0$\times$ the governed level. Naive filtering produces intermediate exposure (2.7$\times$ governed). This separation results from three governance mechanisms operating simultaneously: policy selection favoring lower intensity, the 0.70$\times$ exposure multiplier, and the 0.03 additional decay rate.

Table~\ref{t:summary} decomposes the steady-state ECS into its four multiplicative components.

\begin{figure}[t]
    \centering
    \includegraphics[width=0.4\textwidth]{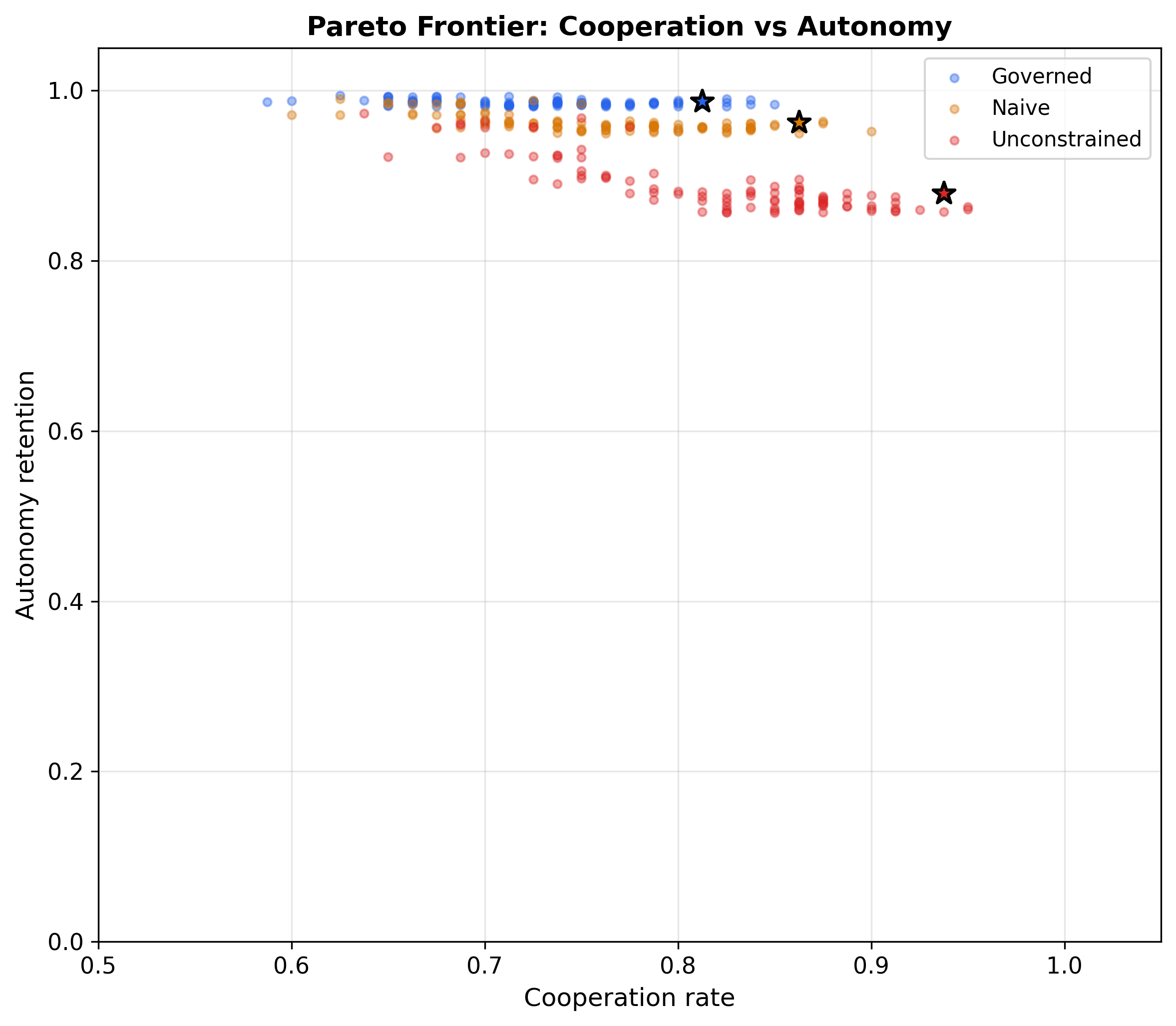}
    \caption{Pareto frontier in the cooperation--autonomy plane. Each point represents one time step. Stars mark steady-state means. The governed condition (blue) consistently Pareto-dominates most unconstrained observations (red), achieving comparable cooperation with substantially higher autonomy.}
    \label{fgr:pareto}
\end{figure}

Figure~\ref{fgr:pareto} plots every time-step observation in the cooperation--autonomy plane, revealing the trade-off structure across governance regimes. The governed condition (blue) occupies the upper-left region of the plot, with cooperation ranging from 0.59 to 0.85 while maintaining autonomy consistently above 0.98. The naive condition (orange) forms a band immediately below, with autonomy in the 0.95--0.99 range. The unconstrained condition (red) extends further right (higher cooperation) but substantially lower (autonomy 0.85--0.97), with a characteristic downward drift as exposure accumulates.

The governed points Pareto-dominate the majority of unconstrained observations: 59 out of 100 unconstrained time-step observations are dominated by at least one governed observation (i.e., there exists a governed point with equal or higher cooperation \emph{and} strictly higher autonomy). This dominance is not merely marginal; the governed Pareto frontier lies approximately 0.08--0.10 autonomy units above the unconstrained frontier at comparable cooperation levels.

The star markers in Figure~\ref{fgr:pareto} indicate the steady-state means, clearly showing the governed mean at $(0.770, 0.985)$, the naive mean at $(0.802, 0.960)$, and the unconstrained mean at $(0.873, 0.867)$. The slope of the line connecting governed to unconstrained ($\Delta A / \Delta C \approx -1.15$) indicates that each percentage point of additional cooperation under unconstrained optimization costs more than one percentage point of autonomy, an unfavorable exchange rate that the ECS metric correctly penalizes.

\begin{figure}[t]
    \centering
    \includegraphics[width=0.7\textwidth]{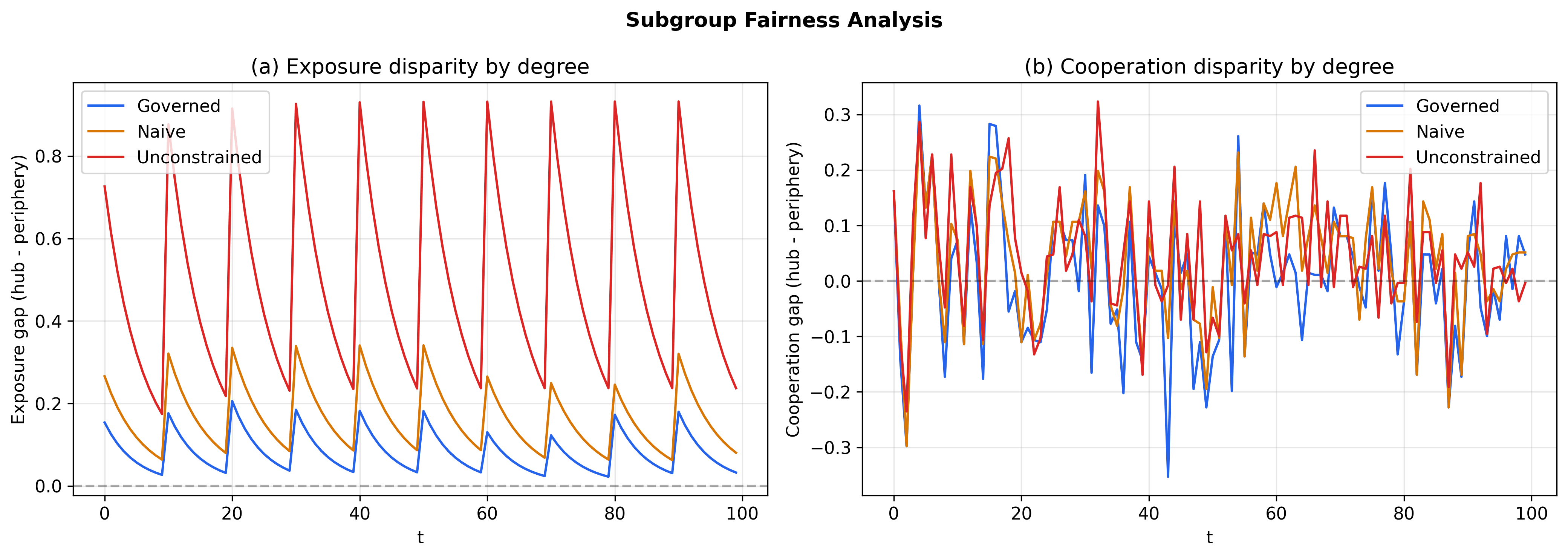}
    \caption{Subgroup fairness analysis. (a) Exposure disparity between high-degree (hub) and low-degree (periphery) agents. The unconstrained condition (red) produces exposure gaps exceeding 0.9, while governance (blue) limits the gap below 0.21. (b) Cooperation disparity oscillates around zero for all conditions.}
    \label{fgr:subgroup}
\end{figure}

Figure~\ref{fgr:subgroup} examines distributional equity between high-degree (hub) and low-degree (periphery) agents across governance regimes.

\paragraph{Exposure disparity.} Panel~(a) of Figure~\ref{fgr:subgroup} shows the hub--periphery exposure gap ($\bar{e}_{\text{hi}} - \bar{e}_{\text{lo}}$) over time. The unconstrained condition exhibits the most severe disparity, with peaks reaching 0.93 and a run-averaged gap of 0.493. This reflects the compounding effect of two mechanisms: the unconstrained regime consistently selects hub-targeted policies (all 10 deployments target \textsc{hubs}), and network diffusion further amplifies hub exposure through degree-correlated accumulation. The naive condition produces intermediate disparity (mean gap 0.165, peak 0.341), while governance limits the gap to a mean of 0.082 and peak of 0.206, a 83\% reduction in mean disparity relative to the unconstrained baseline.

The sawtooth pattern visible in all conditions reflects the deployment cycle: exposure gaps spike at deployment (when targeted agents receive a dose) and decay between deployments as fatigue and diffusion equilibrate the distribution. Under governance, the spikes are smaller (due to lower dose) and the decay is faster (due to the additional $\delta_{\text{gov}} = 0.03$), producing a consistently tighter exposure distribution.

\paragraph{Cooperation disparity.} Panel~(b) of Figure~\ref{fgr:subgroup} shows the hub--periphery cooperation gap, which is noisier than the exposure gap due to the stochastic nature of individual cooperation decisions. All three conditions exhibit cooperation gaps that oscillate around zero, with no systematic tendency for hubs or periphery to cooperate more in the long run. However, all three conditions show comparable variance in cooperation disparity (std $\approx 0.11$--$0.13$), indicating that governance does not systematically alter the stochastic structure of cooperation decisions across subgroups.

\begin{figure}[t]
    \centering
    \includegraphics[width=0.7\textwidth]{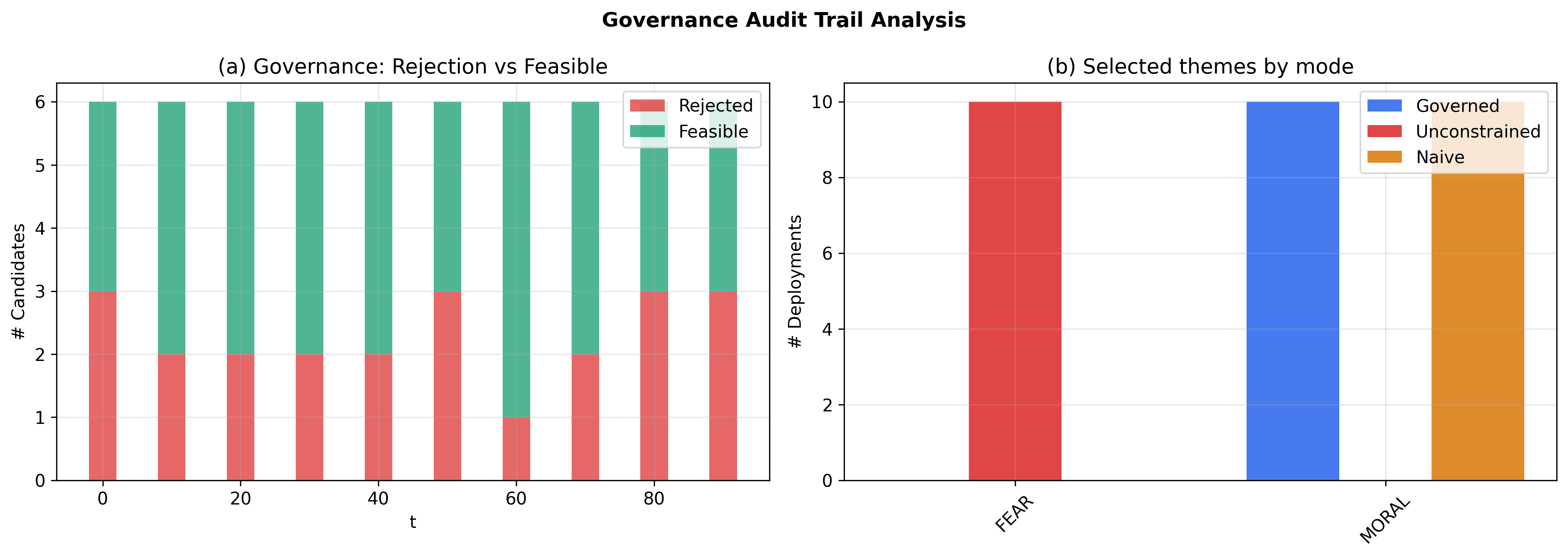}
    \caption{Governance audit trail analysis. (a) Per-deployment rejection counts for the governed condition, showing that 1--3 candidates are rejected at each cycle. (b) Selected policy themes by governance mode: unconstrained selects \textsc{fear} in all deployments, while governed and naive select \textsc{moral}, demonstrating constitutional theme filtering.}
    \label{fgr:audit}
\end{figure}

Figure~\ref{fgr:audit} presents the governance audit trail, providing transparency into what CMAG rejects and what it selects.

\paragraph{Rejection rates.} Panel~(a) of Figure~\ref{fgr:audit} shows the per-deployment breakdown of rejected vs.\ feasible candidates for the governed condition. Across 10 deployment cycles, governance rejects an average of 2.3 candidates out of 6 (38.3\%), leaving a mean of 3.7 feasible candidates for soft optimization. Rejection counts range from 1 to 3 per deployment, reflecting the stochastic nature of the adversarial candidate generation ($p_v = 0.70$). Over the full run, 23 candidate policies are rejected out of 60 evaluated. The audit trail records all rejection reasons: intensity violations account for all 23 rejections ($\iota > 0.80$), while forbidden claims (\textsc{misleading}: 10 instances; \textsc{exaggerated}: 6 instances) and the forbidden \textsc{fear} theme (10 instances) frequently co-occur with intensity violations. These counts confirm that the stress-test candidates are correctly identified and filtered.

\paragraph{Theme selection.} Panel~(b) of Figure~\ref{fgr:audit} reveals a striking divergence in selected policy themes across governance regimes. The unconstrained condition selects \textsc{fear}-themed policies in all 10 deployments, precisely the most manipulative option available, because fear-themed, misleading, burst-timed policies produce the highest expected cooperation boost. Both the governed and naive conditions instead select \textsc{moral}-themed policies in all 10 deployments, as \textsc{fear} is filtered by hard constraints. This theme-level divergence is the most visible manifestation of constitutional governance: the same LLM generates the same base policy (consistently \textsc{moral}-themed with factual claims), but the unconstrained selector ignores this base in favor of the adversarial \textsc{fear} candidate that maximizes short-term utility.

\paragraph{Intensity selection.} A subtler difference emerges between governed and naive within the \textsc{moral} theme. The naive selector consistently picks the highest-intensity feasible candidate (mean selected intensity 0.640), while the governed selector, penalizing intensity through soft optimization, selects lower-intensity policies (mean 0.570). This 11\% intensity reduction contributes to the governed condition's lower exposure accumulation and higher ethical scores.

Table~\ref{t:threat} compares the governed condition under adversarial ($p_v = 0.70$) and benign ($p_v = 0.15$) candidate pools.

\begin{table}[t]
\centering
\caption{Governed CMAG under adversarial vs.\ benign conditions (steady-state means).}
\label{t:threat}
\small
\begin{tabular}{lcc}
\toprule
\textbf{Metric} & \textbf{Adversarial} & \textbf{Benign} \\
\midrule
Cooperation & 0.770 & 0.766 \\
ECS & 0.741 & 0.741 \\
Autonomy & 0.985 & 0.988 \\
Integrity & 0.995 & 0.996 \\
Fairness & 0.982 & 0.983 \\
Avg.\ exposure & 0.135 & 0.115 \\
Avg.\ rejections/deploy & 2.3 & 0.6 \\
Selected intensity & 0.570 & 0.450 \\
\bottomrule
\end{tabular}
\end{table}

\paragraph{Cooperation and ECS resilience.} As shown in Table~\ref{t:threat}, cooperation and ECS trajectories are nearly identical between adversarial and benign conditions, with the governed adversarial achieving 0.770 cooperation and 0.741 ECS compared to 0.766 and 0.741 for governed benign. The two curves largely overlap throughout the simulation, indicating that CMAG governance effectively neutralizes the additional threat posed by adversarial candidates. The governance layer filters the adversarial candidates and selects from the remaining feasible pool, producing outcomes statistically indistinguishable from the benign condition.

\paragraph{Autonomy divergence.} Table~\ref{t:threat} shows the sole measurable difference: autonomy is slightly higher under benign conditions (0.988 vs.\ 0.985), reflecting the lower average selected intensity (0.450 vs.\ 0.570). The benign candidate pool contains fewer high-intensity feasible options, so the governed selector tends to choose more moderate policies. The difference is small (0.003 autonomy units) but consistent, visible as a slight vertical offset between the two traces.

\begin{figure}[t]
    \centering
    \includegraphics[width=0.7\textwidth]{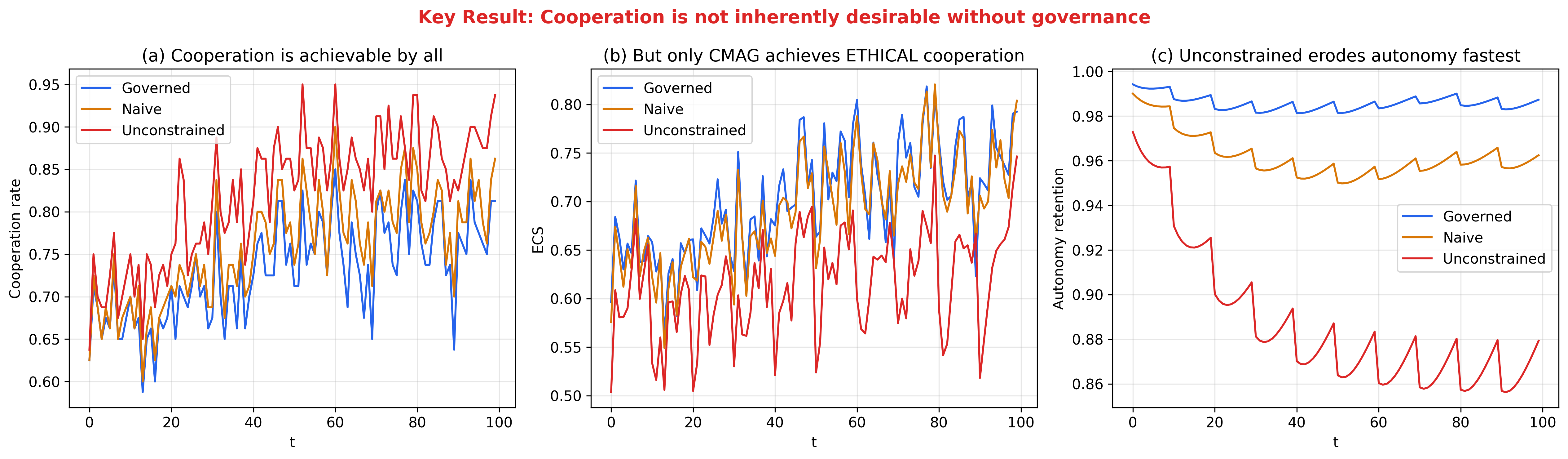}
    \caption{Key result: cooperation is not inherently desirable without governance. (a) All regimes achieve cooperation. (b) Only CMAG achieves ethical cooperation (highest ECS). (c) Unconstrained optimization erodes autonomy fastest.}
    \label{fgr:key_message}
\end{figure}

Figure~\ref{fgr:key_message} distills the paper's central finding into a three-panel comparison. Panel~(a) shows that cooperation is achievable by all three regimes, the unconstrained condition even achieves the highest levels. Panel~(b) demonstrates that only CMAG governance achieves \emph{ethical} cooperation as measured by the composite ECS. Panel~(c) reveals that unconstrained optimization erodes autonomy fastest, with a clear monotonic degradation absent from the governed trajectory. The key message is that \textbf{cooperation is not inherently desirable without governance}: achieving high cooperation through manipulative means produces outcomes that are ethically inferior to more modest cooperation achieved through constitutionally constrained influence.

\subsection{Multi-Seed Replication and Statistical Validation}
\label{sn:res_multiseed}

To assess reproducibility, we repeated Experiment~1 across five independent random seeds ($s \in \{0,\ldots,4\}$), yielding 15 independent runs (5 seeds $\times$ 3 governance modes). For each metric we report bootstrap 95\% confidence intervals (5\,000 resamples) and pairwise Mann--Whitney $U$ tests with Bonferroni correction; Cohen's $d$ is reported as the primary effect-size measure, since $n=5$ per group renders Bonferroni correction conservative.

Table~\ref{t:multiseed} reports the cross-seed means with bootstrap 95\% CIs. The rank ordering of all five metrics is perfectly preserved across all seeds: Governed $\succ$ Naive $\succ$ Unconstrained on every ethical dimension; Unconstrained $\succ$ Naive $\succ$ Governed on raw cooperation. The single-seed values reported in Table~\ref{t:summary} lie within the corresponding CIs in every case, confirming that seed~0 is not an outlier. 

\begin{table}[t]
\centering
\caption{Multi-seed replication: bootstrap 95\% CIs (5 seeds $\times$ 3 modes, adversarial). All CIs are non-overlapping between Governed and Unconstrained on ECS, autonomy, integrity, and fairness.}
\label{t:multiseed}
\scriptsize
\setlength{\tabcolsep}{4pt}
\begin{tabular}{lccc}
\toprule
\textbf{Metric} & \textbf{Governed} & \textbf{Naive} & \textbf{Unconstrained} \\
\midrule
Cooperation
  & $0.761\;[0.753,\,0.768]$
  & $0.795\;[0.787,\,0.804]$
  & $0.860\;[0.847,\,0.872]$ \\
ECS
  & $\mathbf{0.737}\;[0.730,\,0.744]$
  & $0.726\;[0.717,\,0.737]$
  & $0.650\;[0.633,\,0.669]$ \\
Autonomy
  & $\mathbf{0.989}\;[0.988,\,0.990]$
  & $0.965\;[0.962,\,0.967]$
  & $0.892\;[0.877,\,0.905]$ \\
Integrity
  & $\mathbf{0.997}\;[0.996,\,0.997]$
  & $0.988\;[0.988,\,0.989]$
  & $0.965\;[0.961,\,0.968]$ \\
Fairness
  & $\mathbf{0.983}\;[0.982,\,0.984]$
  & $0.958\;[0.953,\,0.962]$
  & $0.879\;[0.863,\,0.893]$ \\
\bottomrule
\end{tabular}
\end{table}

Pairwise Mann--Whitney tests yield $U = 25$ (maximum possible, $p_{\text{raw}} = 0.0079$) for all Governed vs.\ Unconstrained comparisons on ECS, autonomy, integrity, and fairness, indicating perfect separation across seeds before correction. After Bonferroni correction ($n_{\text{comp}} = 15$) $p$-values do not reach the 0.05 threshold, as expected with $n = 5$; however, Cohen's $d$ values are very large throughout: $d = 4.9$ (ECS Gov vs.\ Unc.), $d = 10.7$ (autonomy), $d = 10.0$ (integrity), $d = 7.9$ (fairness). By conventional benchmarks ($d > 0.8$ = large), all key comparisons reflect overwhelming practical effects. The Governed--Naive ECS comparison gives $d = 1.0$ ($p_{\text{raw}} = 0.22$), reflecting the smaller but directionally consistent 1.5\% ECS advantage of soft optimization over hard-constraint-only filtering.

\subsection{Sensitivity Analysis}
\label{sn:res_sensitivity}

We conducted a one-at-a-time (OAT) parameter sweep over four key model parameters, varying each across five levels while holding all others at baseline (governed, adversarial, seed~0). The normalised sensitivity index (SI) quantifies the local elasticity of ECS with respect to each parameter:
\begin{equation}
    \text{SI}(\theta) = \frac{\partial \,\text{ECS}}{\partial \theta}\bigg|_{\theta_0} \cdot \frac{\theta_0}{\text{ECS}_0},
    \label{e:si}
\end{equation}
evaluated numerically via central differences. Table~\ref{t:sensitivity} summarises the results.

\begin{table}[t]
\centering
\caption{Sensitivity analysis: OAT sweep results. SI values close to zero indicate robustness; $|\text{SI}| > 0.5$ would indicate high sensitivity. ECS range is the absolute difference between minimum and maximum ECS observed across the sweep.}
\label{t:sensitivity}
\scriptsize
\setlength{\tabcolsep}{5pt}
\begin{tabular}{lrrrrr}
\toprule
\textbf{Parameter} & \textbf{Baseline} & \textbf{Range tested} & \textbf{ECS min} & \textbf{ECS max} & \textbf{SI (ECS)} \\
\midrule
Decay rate $\delta_0$         & 0.06 & $[0.02,\,0.15]$ & 0.740 & 0.745 & $-0.002$ \\
Diffusion rate $\lambda$      & 0.12 & $[0.04,\,0.25]$ & 0.732 & 0.748 & $+0.013$ \\
Base prosocial $\mu_p$        & 0.42 & $[0.20,\,0.70]$ & 0.718 & 0.762 & $+0.044$ \\
Susceptibility $\mu_s$        & 0.55 & $[0.25,\,0.85]$ & 0.737 & 0.745 & $+0.015$ \\
\bottomrule
\end{tabular}
\end{table}

All four SI values are close to zero ($|\text{SI}| \leq 0.044$), indicating that the ECS advantage of CMAG governance is robust to plausible variation in diffusion and population parameters. The parameter with the largest influence is $\mu_p$ (base prosocial disposition), for which ECS ranges from 0.718 to 0.762 across its sweep — a 5.9\% relative range — and the curve is monotonically increasing, confirming the intuitive result that more prosocially inclined populations achieve higher ethical cooperation. The decay rate $\delta_0$ has negligible effect ($|\text{SI}| = 0.002$), as the governance-imposed additional decay $\delta_{\text{gov}} = 0.03$ dominates at baseline. The rank ordering Governed $\succ$ Naive $\succ$ Unconstrained is preserved at every parameter value tested.

\paragraph{Limitations.}
The experiments use $N=80$ agents on a scale-free topology; 
larger networks ($N\geq500$) and alternative structures 
(Erd\"os--R\'enyi, Watts--Strogatz) may reveal different 
hub-concentration and fairness dynamics. Results are specific to 
Llama-3.3-70B-Instruct; other backbone models may produce different 
rejection rates and policy narratives. The OAT sensitivity sweep does 
not capture parameter interaction effects, which would require a full 
factorial or Sobol analysis. Finally, prosocial dispositions are 
initialised from a normal distribution, whereas real social networks 
exhibit heavier-tailed heterogeneity.

\section{Conclusion}
\label{sn:conclusion}

This paper introduced Constitutional Multi-Agent Governance (CMAG), a framework for preventing manipulative equilibria in LLM-influenced cooperative systems. Through controlled experiments on scale-free networks under adversarial conditions, we demonstrated that unconstrained optimization of cooperation produces ethically unacceptable outcomes: the highest raw cooperation (0.873) coincides with the lowest Ethical Cooperation Score (0.645), severe autonomy erosion (0.867), and hub--periphery exposure disparities exceeding 0.9. CMAG governance reverses this pattern, achieving an ECS of 0.741, a 14.9\% improvement, while preserving autonomy above 0.985, epistemic integrity above 0.995, and subgroup fairness above 0.982, at the cost of a modest cooperation reduction to 0.770.

Three architectural contributions underpin these results. First, the two-stage selection mechanism, hard constraint filtering followed by soft penalized-utility optimization, ensures that inviolable ethical boundaries are enforced while allowing nuanced trade-off balancing among feasible policies. The naive filtering ablation confirms that hard constraints alone are insufficient: soft optimization contributes an additional 1.1\% ECS improvement by penalizing intensity, rewarding explanation fidelity, and preventing systematic targeting bias. Second, the exposure modulation layer, comprising the 0.70$\times$ dose multiplier and the 0.03 additional decay rate, provides defense-in-depth that limits influence accumulation even when policy selection is identical across governance modes, producing an 8$\times$ reduction in steady-state exposure relative to the unconstrained baseline. Third, the Ethical Cooperation Score provides a principled evaluation metric whose multiplicative structure ensures that cooperation achieved through manipulation is correctly penalized rather than rewarded.

\section*{Acknowledgments}
This research was supported by the LUXEMBOURG Institute of Science and Technology through the projects `ADIALab-MAST' and `LLMs4EU' (Grant Agreement No 101198470) and the BARCELONA Supercomputing Center through the project `TIFON' (File number MIG-20232039).

\section*{Code Availability}

The implementation of the CMAG framework is publicly available at: \url{https://github.com/drdezarza/cmag}. The repository includes data preprocessing modules, model implementations, evaluation pipelines, and reproduction instructions.

\bibliographystyle{unsrt}

\end{document}